\documentclass{iopart}
\usepackage{iopams} 
\usepackage{mathptmx}
\usepackage[utf8]{inputenc} 
\usepackage{graphicx}
\usepackage{epstopdf} 
\usepackage[numbers,comma,square,sort&compress,merge]{natbib} 
\usepackage{xspace}
\usepackage[
,textwidth=17.5cm
,textheight=23.0cm 
,verbose
,dvips
]{geometry}
\usepackage{url}

\renewenvironment{indented}{\begin{indented}}{\end{indented}}

\usepackage{multicol}

\begin{document}

\topical{Luminescence associated with stacking faults in GaN}

\author{Jonas Lähnemann, Uwe Jahn, Oliver Brandt, Timur Flissikowski, Pinar Dogan and Holger T. Grahn}
\address{Paul-Drude-Institut für Festkörperelektronik, Hausvogteiplatz 5--7, 10117 Berlin, Germany}
\ead{laehnemann@pdi-berlin.de}

\begin{abstract} 
Basal-plane stacking faults are an important class of optically active structural defects in wurtzite semiconductors. The local deviation from the 2H stacking of the wurtzite matrix to a 3C zinc-blende stacking induces a bound state in the gap of the host crystal, resulting in the localization of excitons. Due to the two-dimensional nature of these planar defects, stacking faults act as quantum wells, giving rise to radiative transitions of excitons with characteristic energies. Luminescence spectroscopy is thus capable of detecting even a single stacking fault in an otherwise perfect wurtzite crystal. This review draws a comprehensive picture of the luminescence properties related to stacking faults in GaN. The emission energies associated with different types of stacking faults as well as factors that can shift these energies are discussed. In this context, the importance of the quantum-confined Stark effect in these zinc-blende/wurtzite heterostructures, which results from the spontaneous polarization of wurtzite GaN, is underlined. This discussion is extended to zinc-blende segments in a wurtzite matrix. Furthermore, other factors affecting the emission energy and linewidth of stacking fault-related peaks as well as results obtained at room temperature are addressed. The considerations presented in this article should be transferable also to other wurtzite semiconductors.
\end{abstract}

\pacs{78.55.Cr,
78.60.Hk,
61.72.Nn,
81.07.St
}

\submitto{\JPD}

\twocolumn


\section{Introduction}

GaN is the most prominent semiconductor used for applications in solid state lighting~\cite{Pimputkar_natphot_2009}. To improve the efficiency of GaN-based devices, various strategies are employed with the aim to reduce the high dislocation densities inherent to the heteroepitaxial growth of GaN and avoiding polarization fields which are detrimental to optical devices ~\cite{Waltereit_nat_2000,Speck_mrsbull_2009}. However, these growth strategies often result in the formation of stacking faults: Especially the growth of non- and semi-polar GaN layers \cite{Craven_apl_2002,Sun_jap_2002,Paskov_jap_2005,Baker_jjap_2006, Vennegues_jap_2012} as well as epitaxial-lateral overgrowth \cite{Haskell_apl_2003,Haskell_apl_2005,Liu_apl_2005,Guhne_jap_2007, Bastek_apl_2008,Zhu_jap_2010} or the coalescence overgrowth of nanowires \cite{Bougrioua_jcg_2007,Cherns_apl_2008,Dogan_jcg_2011,Korona_jl_2014} are associated with the formation of basal-plane stacking faults independent of which growth technique is used. As a consequence, the emission characteristics and transport properties of the layers are changed \cite{Liu_apl_2005,Konar_apl_2011}.

Basal-plane stacking faults (BSFs) in GaN are a local deviation from the hexagonal (2H) wurtzite (WZ) to the cubic (3C) zinc-blende (ZB) crystal structure. However, standard methods for the structural characterization of such crystals (e.g. X-ray diffraction) are not sensitive enough to detect low densities of BSFs. Only transmission electron microscopy, a rather costly and time-consuming method, facilitates the clear identification of BSFs. Alternatively, luminescence spectroscopy can be a sensitive tool to detect the presence of BSFs in semiconductor thin films or nanostructures \cite{Paskov_pssc_2008}, because BSFs are optically active defects. As first pointed out by Rieger \etal~\cite{Rieger_prb_1996} and Rebane \etal~\cite{Rebane_pssa_1997}, these WZ/ZB heterostructures may be pictured as ZB quantum wells (QWs) in a WZ matrix leading to emission at energies lower than for excitons in the bulk WZ phase, which in GaN is found at 3.478~eV \cite{Korona_prb_2002}. While transitions associated with excitons bound to BSFs of the intrinsic I$_1$ type are well established to lie in the range of 3.40 to 3.42~eV~\cite{Paskov_jap_2005,Liu_apl_2005,Albrecht_mrssp_1997, Fischer_jcg_1998,Salviati_pssa_1999,Leroux_jap_1999,Mah_jcg_2001, Skromme_MaterialsScienceForum_2004}, there are fewer reports on luminescence lines associated with BSFs of the intrinsic I$_2$ type~\cite{Sun_jap_2002,Paskova_procspie_2006,Tischer_prb_2011, Lahnemann_prb_2012}. Only recently, we reported emission peaks related to extrinsic BSFs \cite{Lahnemann_prb_2012}, and Jacopin \etal~\cite{Jacopin_jap_2011} observed luminescence lines associated with ZB segments.

However, BSFs and polytypism may not always be undesirable. For the classic III-V semiconductors, even the growth of controlled heterostructures of different crystal polytypes has been reported, and the possibility to use these for device applications has been brought up \cite{Caroff_natnano_2009,Dick_nl_2010}. Furthermore, stacking fault-related luminescence can be utilized as a probe for crystal properties \cite{Spirkoska_prb_2009,Jahn_prb_2012,Graham_prb_2013,Lahnemann_prb_2012, Nogues_apl_2014}. Specifically, polytypic GaAs nanowires have been used to investigate the metastable WZ phase \cite{Spirkoska_prb_2009,Jahn_prb_2012,Graham_prb_2013}. Recently, we have also used the emission energies associated with BSFs to derive an experimental value for the spontaneous polarization of wurtzite GaN \cite{Lahnemann_prb_2012}. The combination of this polarized polytype with the unpolarized ZB phase presents a unique system to address the spontaneous polarization independent from the piezoelectric polarization. Also, the diffusion of carriers to a BSF evidenced by the intensity distribution in cathodoluminescence images has been used to derive the diffusion length in GaN nanowires \cite{Nogues_apl_2014}.

In this article, we present a review on the luminescence properties related to stacking faults in GaN exemplified by some of our experimental data. We give a comprehensive overview of a number of aspects mentioned in the literature, which can explain the varying results observed in different studies. As we will point out, the physics of BSFs includes some factors which require an extension of the simple QW picture. The impact of the spontaneous polarization as well as a coupling to point defects and other BSFs need to be considered. To start with, section~\ref{sec:sfs} gives an introduction to stacking faults in WZ crystals and discusses why they can be considered as ideal QWs. Section~\ref{sec:lumi} begins with a review of luminescence energies related to stacking faults observed in the literature illustrated with experimental results. The importance of the spontaneous polarization for the luminescence related to BSFs and its manifestation in the emission characteristics are discussed in section~\ref{sec:psp}. 
Then, section~\ref{sec:othershifts} introduces other factors that affect the emission energy related to such WZ/ZB quantum well structures, while section~6 presents measurements of BSF related luminescence lines at room temperature. Both cathodoluminescence (CL) with a high spatial resolution ($\approx 20$~nm) and micro-photoluminescence ($\mu$PL) spectroscopy are employed in the course of this work. Finally, section~\ref{sec:concl} summarizes and concludes this work. In the appendix, experimental aspects such as the growth of the investigated samples and details of the instruments used for luminescence spectroscopy are given.

\section{Stacking faults in GaN}\label{sec:sfs}

%
\begin{figure}
\centering
\includegraphics*[width=6cm]{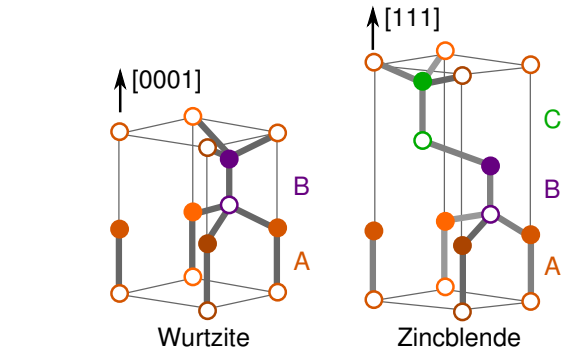}
\caption{\label{fig:struct}Primitive wurtzite and non-primitive zinc-blende unit cells. Filled circles denote anions, while open circles denote cations. The stacking sequence of the layers is indicated by the different colours and the letters on the side.}
\end{figure}

In equilibrium, GaN crystallizes in the wurtzite (WZ) phase, a \emph{hexagonally close-packed} (hcp) crystal structure with two different sub-lattices. However, also the closely related metastable \emph{cubic close-packed} (ccp) zinc-blende (ZB) structure can be synthesized by epitaxy on suitable substrates~\cite{Paisley_jvsta_1989,Strite_jvstb_1991,Menniger_prb_1996,Renard_apl_2010}.

Both crystal structures exhibit tetrahedral coordination and the same nearest neighbour configuration. However, they differ in the bond angle of their third-nearest neighbour configuration. In other words, their close-packed stacking sequence differs as depicted in figure~\ref{fig:struct}. The WZ stacking sequence along the (0001) direction is denoted as $a\alpha b\beta a\alpha b\beta\dots$, while ZB exhibits an $a\alpha b\beta c\gamma a\alpha b\beta c\gamma\dots$ stacking along the (111) direction, where the Latin and Greek letters refer to cations and anions, respectively. As a pair of cations and anions always shares the same in-plane position, the simplified terminology of an \emph{ABAB\dots} stacking for WZ and \emph{ABCABC\dots} for ZB is often used. The ZB structure can also be seen as a face-centred cubic (fcc) structure with a two-atom base. Therefore, ZB has four equivalent polar axes in the (111) directions, while WZ has a singular polar axis in the (0001) direction.

%
\begin{figure}
\centering
\includegraphics*[width=8cm]{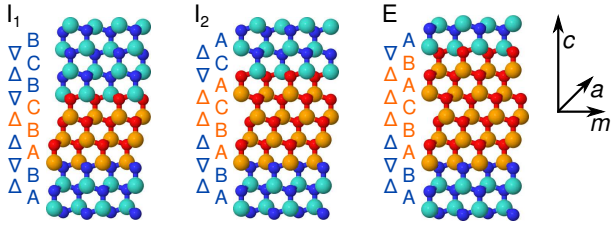}
\caption{\label{fig:bsfs}Stacking sequences for the intrinsic I$_1$ and I$_2$ as well as the extrinsic E stacking fault. The layers for which the ZB stacking sequence is upheld are highlighted, while in the operator notation the breaches to the wurtzite stacking rule are highlighted.}
\end{figure}
 
Stacking faults are local changes in the stacking sequence. In WZ GaN, they thus constitute ZB segments in a WZ matrix. Three types of BSFs are distinguished in the literature \cite{Frank_pm_1953,Chadderton_pm_1963,Blank_pssb_1964,Hirth_1982}, where intrinsic BSFs are formed by the change from one hcp lattice to another one, i.e.\ from \emph{AB} to \emph{BC} or \emph{AC} stacking, while extrinsic BSFs are formed through the insertion of an extrinsic layer, i.e.\ a \emph{C} layer in an \emph{AB} sequence:

\emph{(a) intrinsic I$_1$ BSFs} involve one breach of the WZ stacking rule where the third-nearest neighbour should be the same. They exhibit a stacking sequence of \emph{ABABCBCBC\dots}

\emph{(b) intrinsic I$_2$ BSFs} involve two breaches of the WZ stacking and have the stacking sequence \emph{ABABCACAC\dots}

\emph{(c) extrinsic E BSFs} involve three breaches of the WZ stacking and have the stacking sequence \emph{ABABCABAB\dots}

The stacking sequences related to these three types of BSFs are illustrated in figure~\ref{fig:bsfs}. Additionally to the in-plane positional labels, the operator notation \cite{Frank_pm_1953} is given: The successions $A\to B$, $B\to C$ or $C\to A$ are marked by $\vartriangle$, while their opposites are marked by $\triangledown$. In this notation, the WZ stacking is an alternation $\vartriangle$\,$\triangledown$, while the ZB stacking is represented by a sequence of the same operator. The BSFs can then be characterized by the number of breaches of the WZ stacking rule as (1$\vartriangle$), (2$\vartriangle$) and (3$\vartriangle$)~\cite{Frank_pm_1953}.

These three types of BSFs are clearly defined structural defects. Note that when the ZB stacking sequence is continued for a larger number of layers, we will refer to these structures as \emph{ZB segments}.

Calculations by Stampfl and Van de Walle \cite{Stampfl_prb_1998} confirm that the formation energy of the different BSF types increases in the presented order, which was already inferred by Frank and Nicholas \cite{Frank_pm_1953} from the increasing number of breaches of the WZ stacking. In contrast to the other two types, I$_2$ BSFs can be formed by a single slip in the basal plane \cite{Frank_pm_1953} and may therefore be introduced by post-growth deformation. When ending within a crystal, BSFs must be terminated by dislocations. Specifically, these are partial dislocations with Burger's vector $\mathbf{b}=\frac{1}{6}[20\bar{2}3]$ (Frank-type) in the case of I$_1$ BSFs, with $\mathbf{b}=\frac{1}{3}[10\bar{1}0]$ (Shockley-type) in the case of I$_2$ BSFs, and with $\mathbf{b}=\frac{1}{2}[0001]$ (Frank-type) in the case of E BSFs~\cite{Drum_pm_1965,Hirth_1982}. Otherwise, BSFs can be terminated by free surfaces or heteroepitaxial interfaces (in the case of non- or semi-polar growth). I$_1$ BSFs are the most common type observed, as they have the lowest formation energy. For non-polar growth directions, it has been shown that the formation of I$_1$ BSFs may be induced by the coalescence of three-dimensional islands in the Volmer-Weber growth mode~\cite{Trampert_jpiv_2006,Vennegues_jap_2012}. However, also I$_2$ BSFs are commonly observed in non-polar samples~\cite{Sun_jap_2002,Liu_pml_2004,Paskova_procspie_2006}. Their nucleation might for example become favourable at atomic steps at the substrate interface. The probability of E BSFs or thicker cubic segments being formed may increase when free surfaces are available, such as for the growth of nanowires.

Additionally to the (0001) basal planes, stacking faults can occur on prismatic ($11\bar{2}0$)-planes (\emph{A}-planes) of WZ crystals~\cite{Blank_pssb_1964,Drum_pm_1965}. These prismatic stacking faults (PSFs) exhibit a displacement vector of $\mathbf{b}=\frac{1}{2}[10\bar{1}1]$ and always connect two I$_1$ BSFs with stair-rod dislocations forming at the intersection~\cite{Drum_pm_1965}. This folding of stacking faults can lead to step-like arrangements and even loops~\cite{Blank_pssb_1964}. Only the tetrahedral bond angles in the prismatic plane are changed in the creation of PSFs so that again no dangling bonds are formed~\cite{Drum_pm_1965}. Their atomic configuration can be seen in transmission electron micrographs of (0001) cross sections~\cite{Hu_apl_2012}.

\subsection{Stacking faults as quantum wells}

%
\begin{figure}
\centering
\includegraphics*[width=6cm]{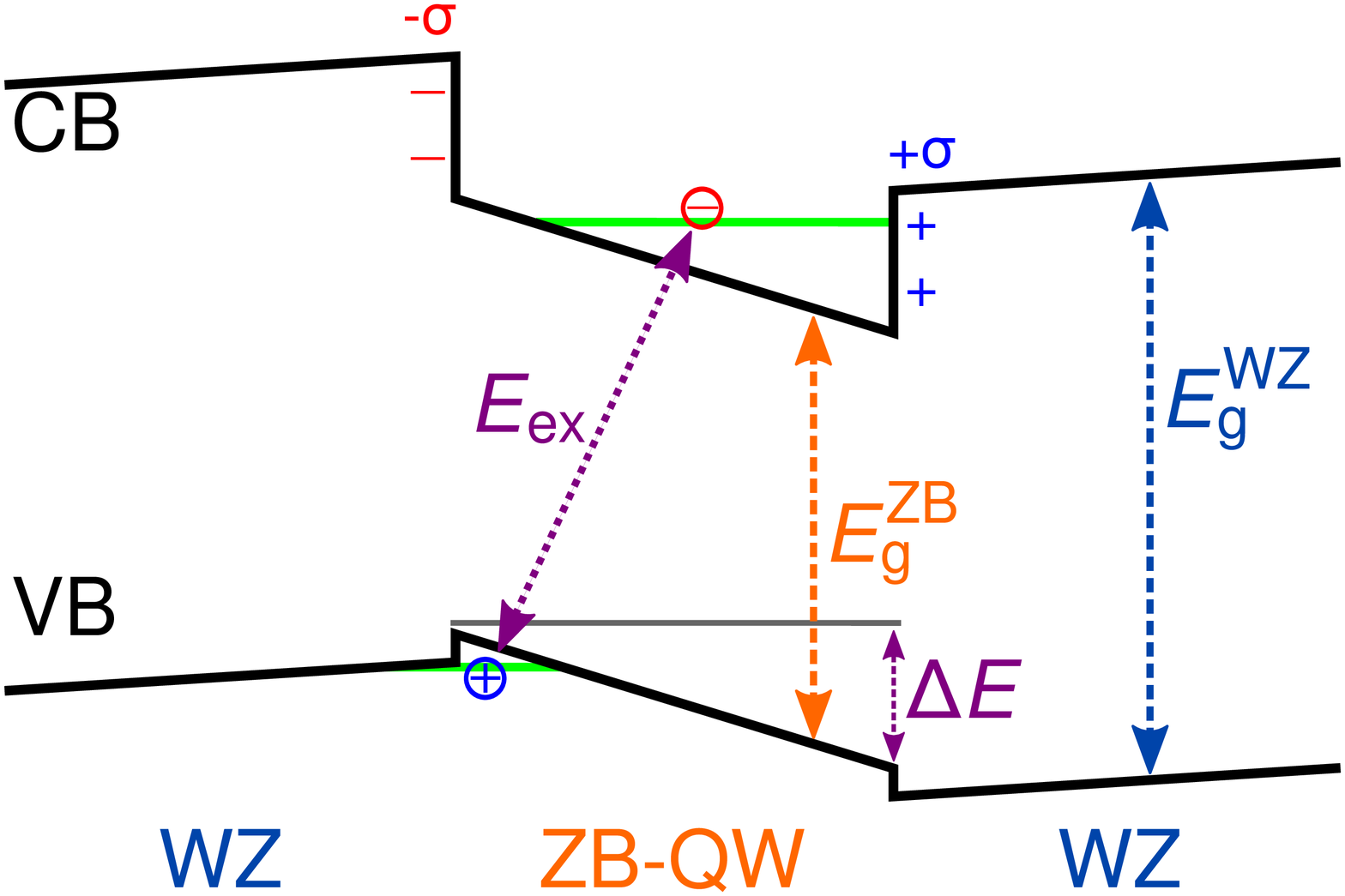}
\caption{\label{fig:band-scheme}Schematic band profile of a stacking fault for a type-I band alignment. The polarization induced sheet charges $\sigma$ leading to the QCSE and the resulting change in the transition energy $E_{ex}$ by $\Delta E$ are indicated.}
\end{figure}

The idea to picture BSFs as QWs was first suggested by Rieger \etal~\cite{Rieger_prb_1996} and Rebane \etal~\cite{Rebane_pssa_1997}. With the emission energy of free excitons being 3.478~eV for WZ GaN \cite{Korona_prb_2002} and 3.276~eV for ZB GaN~\cite{Menniger_prb_1996,Renard_apl_2010}, the difference in band gap between these two polytypes of GaN amounts to 202~meV. Excitons can bind to these structural defects, and BSFs can thus be considered as \emph{ideal} QWs with atomically flat interfaces, negligible lattice mismatch ($\Delta a/a < 2 \times 10^{-3}$ \cite{Brandt_1998}), and without alloy fluctuations. However, as a result of its singular polar axis, the WZ phase exhibits a spontaneous polarization $P_\mathrm{sp}$. The discontinuity of this spontaneous polarization at the WZ/ZB interface results in a charged layer at the interface and thus leads to an electric field in the QW. As a consequence, the emission energy is redshifted by what is known as the \emph{quantum-confined Stark effect} (QCSE)~\cite{Miller_prl_1984,Takeuchi_jjap_1997}.
The importance of the polarization fields was first pointed out by Sun \etal \cite{Sun_jap_2002} in their observation of I$_2$ BSFs and by Skromme \etal \cite{Skromme_MaterialsScienceForum_2004}. We have recently seen experimental evidence for the presence of such polarization fields \cite{Lahnemann_prb_2012} as further discussed in section~\ref{sec:psp}. The resulting band profile of a ZB QW in a WZ matrix for a type-I band alignment is sketched in figure~\ref{fig:band-scheme}. 

At this point, the question arises, whether the band alignment for WZ/ZB heterostructures is of type I (offset of conduction and valence band have opposite sign) or type II (staggered band offset). Calculations based on density-functional theory (DFT) disagree on this point with some studies giving a type-I \cite{Majewski_MRSInternetJournalofNitrideSemiconductorResearch_1998,Belabbes_prb_2011}  and others a type-II \cite{Murayama_prb_1994,Stampfl_prb_1998} band alignment. In experimental studies, a blueshift of the QW emission energy with increasing excitation density is sometimes seen as evidence for a type-II band alignment~\cite{Lu_apl_2003,Jacopin_jap_2011}. However, due to the strong spontaneous polarization of WZ GaN, electron and hole will be spatially separated at BSFs independent of the actual band alignment~\cite{Belabbes_prb_2011}, as also indicated by the sketch in figure~\ref{fig:band-scheme}. In fact, the DFT calculations in \cite{Belabbes_prb_2011} give a type-I band alignment, but when calculating the wavefunctions of electrons and holes for a WZ/ZB quantum well, holes are largely confined in the WZ region as for a type-II offset. Therefore, the issue of the band alignment is so far unresolved. 

Another critical point is the definition of the thickness of stacking faults (cf.\ figure~\ref{fig:bsfs}). This question is crucial for band profile calculations of stacking faults acting as quantum wells. A first possibility is to take the number of breaches of the hexagonal stacking rule, i.e.\ layers for which the third-nearest neighbour is not the same (operator notation in figure~\ref{fig:bsfs}). According to this definition, the thickness of the I$_1$, I$_2$ and E BSF would be 1, 2 and 3 bilayers (molecular monolayers) or $0.5c_0$, $c_0$ and $1.5c_0$ (where $c_0$ is the c-lattice constant), respectively. A second definition uses the number of bilayers for which the cubic \emph{ABC} stacking sequence is upheld~\cite{Rebane_pssa_1997}. This definition gives 3, 4 and 5 bilayers (or $1.5c_0$, $2c_0$ and $2.5c_0$) for the three types of BSFs. The latter definition would be intuitive from the structure of the stacking faults seen in figure~\ref{fig:bsfs}. Furthermore, the envelope function formalism used for calculating the eigenvalues of the quantized states in a QW with abrupt interfaces reaches its limits for QWs of only a few atomic layers \cite{Burt_jpcm_1992}. We have resolved these questions in \cite{Lahnemann_prb_2012} by introducing an \emph{effective electronic thickness} determined from the miscroscopic electrostatic potential of the E BSF calculated by DFT. From these calculations, we get 3.7 bilayers as the thickness of the E BSF, and thus 2.7 bilayers for the I$_2$ and 1.7 bilayers for the I$_1$ BSF.

Quantum wells with the \emph{ideal} properties mentioned above should result in a narrow linewidth of the associated luminescence spectra \cite{Akopian_nl_2010,Graham_prb_2013}. However, several factors can lead to shifts in the emission energy for excitons bound to BSFs and thus to a broadening of the observed peaks. In sections~\ref{sec:psp} and~\ref{sec:othershifts}, factors influencing the emission energy and thus possible origins for such energy shifts are discussed. First, however, let us turn to an overview of the observed emission energies related to BSFs.

\section{Luminescence lines associated with stacking faults}\label{sec:lumi}

%
\begin{table}[b]
\centering
\caption{\label{tab:energies}Summary of low-temperature emission energies associated with BSFs, of other common structural or point defects emitting in the same spectral range and of the free exciton (FE) emission. For the former, our recommended values are compared to other values from the literature, which are listed as reported in the literature and might not always be corrected for strain in the structures. All values are given in eV.}
\begin{tabular}{lcl}
\br
 & Our value & Values from literature (eV)\\
\mr
WZ FE & & 3.478~\cite{Korona_prb_2002}\\
ZB FE & & 3.276~\cite{Menniger_prb_1996,Renard_apl_2010}\\
I$_1$ BSFs & 3.42 & 3.40--3.42~\cite{Albrecht_mrssp_1997,Salviati_pssa_1999,Leroux_jap_1999,Mah_jcg_2001,Skromme_MaterialsScienceForum_2004,Liu_apl_2005,Paskov_jap_2005}\\
I$_2$ BSFs & 3.35 & 3.32--3.36~\cite{Tischer_prb_2011,Paskova_procspie_2006,Sun_jap_2002} \\
E BSFs & 3.29 & \\
\multicolumn{3}{l}{ZB segments $\approx3.0$--3.27}\\
\multicolumn{2}{l}{Prismatic SFs} & 3.21 \cite{Bai_jap_2005}, 3.30 \cite{Mei_apl_2006}, 3.33 eV \cite{Liu_apl_2005}\\
\multicolumn{2}{l}{Dislocations} & 3.27 \cite{Bertram_ICNS_2013}, 3.29 \cite{Liu_apl_2005} \\
DAP & & 3.27~\cite{Reshchikov_jap_2005} \\
\br
\end{tabular}
\end{table}

We have recently observed luminescence lines related to all three types of BSFs and reported a consistent set of the associated emission energies~\cite{Lahnemann_prb_2012}. In this study, the emission energies at cryogenic temperatures associated with individual BSFs were deduced from a statistical analysis of peaks related to stacking faults in CL spectral maps. The statistical analysis enabled us to distinguish contributions from individual BSFs from those experiencing an energy shift (possible reasons for such shifts are discussed later). Thereby, we determined low-temperature (10~K) emission energies of 3.42~eV for excitons bound to the I$_1$ type BSFs, 3.35~eV for I$_2$ BSFs and 3.29~eV for E BSFs. The presence of all three types of BSFs in the investigated GaN microcrystals was confirmed by high-resolution transmission electron microscopy (TEM)~\cite{Lahnemann_prb_2012}.

The first experimental assignment of emission lines at 3.40--3.42~eV to I$_1$ BSFs was reported by Albrecht \etal \cite{Albrecht_mrssp_1997} and Salviati \etal \cite{Salviati_msmc_1997,Salviati_pssa_1999} using PL/CL spectroscopy and TEM. This emission energy has been well established in the meantime \cite{Leroux_jap_1999,Mah_jcg_2001,Skromme_MaterialsScienceForum_2004,Liu_apl_2005,Paskov_jap_2005}, particularly through a direct correlation of CL spectroscopy and TEM by Liu \etal~\cite{Liu_apl_2005}.

For I$_2$ BSFs, Sun \etal \cite{Sun_jap_2002} reported emission lines at 3.356~eV and Paskova \etal \cite{Paskova_procspie_2006} at 3.355 eV for \emph{M}- respectively \emph{A}-plane GaN with a high density of this type of BSFs. A direct correlation between CL spectroscopy and high-resolution TEM was carried out in \cite{Tischer_prb_2011}. However, the value of 3.32~eV reported in \cite{Tischer_prb_2011} is redshifted by 0.02~eV through tensile strain as the excitonic near-band edge (NBE) emission in the same region of the sample is found at 3.45~eV (instead of at the strain-free value of 3.47~eV). We conclude that the value of 3.35~eV obtained in our statistical analysis confirms the existing experimental data on the emission energy related with I$_2$ BSFs.

Naturally, there is a variety of other luminescence lines in the spectral range associated with emission from BSFs. Notably (concerning extended defects), Liu \etal~\cite{Liu_apl_2005} found that \emph{A}-plane PSFs connecting BSFs of the I$_1$ type give rise to luminescence lines at around 3.33~eV, while partial dislocations terminating the I$_1$ BSFs are related to emission at 3.29~eV. Often, dislocations are found to act as nonradiative recombination centres \cite{Jahn_prb_2010}. However, the luminescence lines related to partial dislocations might be a result of impurity decoration of the dislocation cores~\cite{Liu_apl_2005}. Similar results were reported by Mei \etal~\cite{Mei_apl_2006} with an emission energy of 3.30~eV related to \emph{A}-plane PSFs, whereas Bai \etal \cite{Bai_jap_2005} assign an emission at 3.21~eV to PSFs and/or associated stair-rod partial dislocations. Bertram \etal~\cite{Bertram_ICNS_2013} used CL in a TEM to link an emission line at 3.27~eV to stair-rod dislocations in GaN. Finally, different point defects or sample contamination can lead to luminescence bands in this spectral range~\cite{Reshchikov_jap_2005}, particularly the donor-acceptor-pair (DAP) transition at 3.27~eV.

%
\begin{figure}
\centering
\includegraphics*[width=8cm]{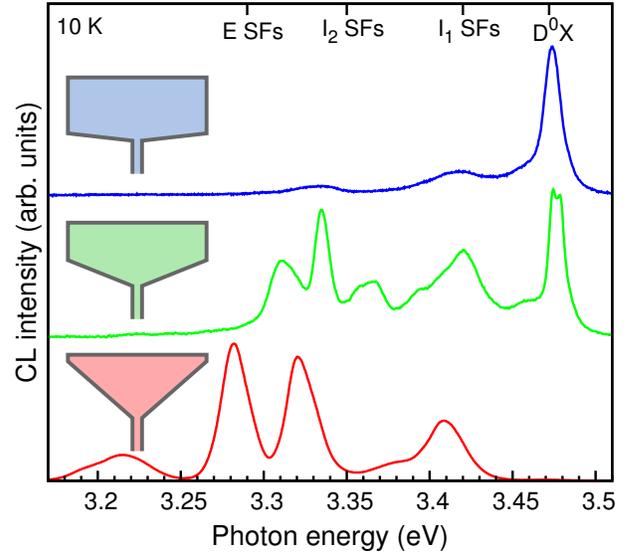}
\caption{\label{fig:cl}Integral CL spectra from different GaN microcrystals with weak (top), intermediate (middle) and strong (bottom) contributions from BSF luminescence. The shapes of the corresponding microcrystals are sketched as insets; the lateral growth of the crystals is associated with the formation of BSFs.}
\end{figure}

The emission energies related to the different types of BSFs are summarized in table~\ref{tab:energies}. For comparison, emission energies associated with other common structural or point defects emitting in this spectral range are also given in the table, and the energies of the free exciton transitions in WZ and ZB GaN are added.

To introduce some of the features of luminescence lines associated with excitons at BSFs (hereafter also referred to as BSF luminescence/emission), typical CL spectra containing contributions from BSF luminescence are displayed in figure~\ref{fig:cl}. The measurements were performed on the cross sections of GaN microcrystals overgrown on GaN nanowires (see appendix for details on the sample). BSFs are formed during the lateral expansion from a single nanowire to the microcrystal [cf.\ figure~\ref{fig:sem}(b)]. While the upper spectrum in figure~\ref{fig:cl} contains only a weak contribution from a few BSFs and is dominated by the NBE luminescence lines, the middle spectrum has significant contributions from both BSFs and the NBE luminescence. The spectrum at the bottom is dominated by BSF emission. These spectral differences are correlated with the shape of the microcrystals as indicated by the sketches given as insets in figure~\ref{fig:cl}. For the bottom spectrum, the faulted region spans the whole cross section of the microcrystal, whereas the other two spectra correspond to microcrystals that exhibit a region free of extended defects towards the surface (cf.\ figure~1 in \cite{Lahnemann_prb_2012}). Each of the microcrystals contains several BSFs of different types, bundles of these or even (thicker) ZB segments, the formation of which seems to be induced by the lateral growth mode.

%
\begin{figure*}
\includegraphics*[width=17cm]{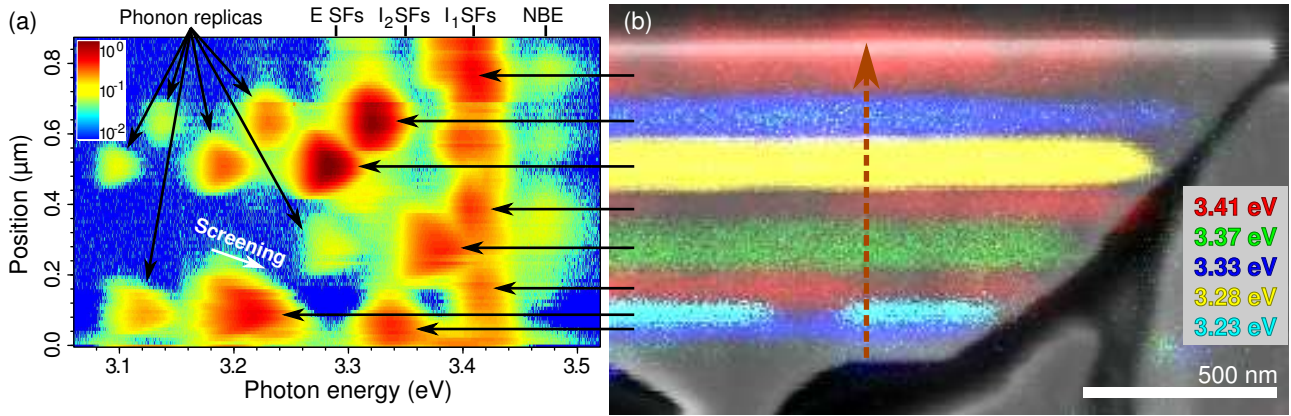} 
\caption{\label{fig:clscan}(a) CL spectral line scan  along the cross section of a GaN microcrystal revealing luminescence lines associated with BSFs and ZB segments of different thickness. The CL intensity is colour-coded on a logarithmic scale. (b) Corresponding SEM image of the GaN microcrystal with the path of the line scan marked by the dashed arrow. Superimposed are false-colour monochromatic CL images recorded at the indicated detection energies. The corresponding emission peaks in the line scan are highlighted by horizontal arrows.}
\end{figure*}

%
\begin{figure}[h]
\centering
\includegraphics*[width=8.5cm]{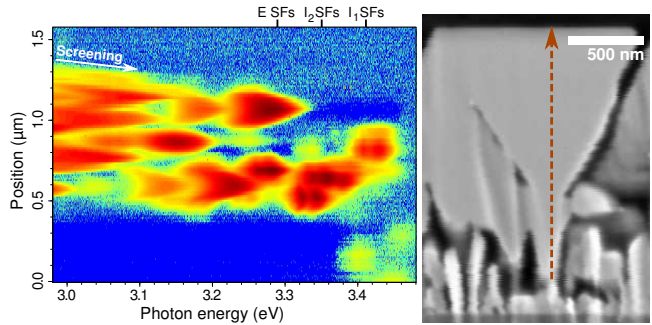} 
\caption{\label{fig:clscan2}CL spectral line scan from the cross section of a second GaN microcrystal and corresponding SEM image with the path of the scan marked by a dashed arrow. The CL intensity is colour-coded on a logarithmic scale. Emission lines ranging down to 3.0~eV, corresponding to zinc-blende segments of about 3~nm thickness, are observed. The blueshift resulting from the field screening under direct CL excitation can amount to about 80--90~meV.}
\end{figure}

Complementary, figures~\ref{fig:clscan} and \ref{fig:clscan2} show CL spectral maps recorded along lines on the cross section of two different microcrystals. The line scan in figure~\ref{fig:clscan}(a) was recorded on the same microcrystal as the bottom spectrum in figure~\ref{fig:cl}. Figure~\ref{fig:clscan}(b) presents monochromatic CL images of that cross section superimposed on an SEM image. These measurements demonstrate that the various peaks observed in the overview spectra for a single microcrystal originate from different positions along the cross section of the crystal, i.e.\ from different BSF quantum wells. In the cross-sectional CL images (non-polar plane), the emission related to BSFs or possibly bundles of these shows a characteristic distribution along the basal plane. The corresponding emission peaks in the spectral map are indicated by horizontal arrows. The broadening along the \emph{c}-axis is the result of the scattering of incident electrons combined with carrier diffusion \cite{Nogues_apl_2014}, while the crescent shape of these emission peaks in the spectral map is the result of an excitation density-dependent screening of the polarization fields, which will be discussed in the next section. As for the integral spectrum (bottom spectrum in figure~\ref{fig:cl}), hardly any NBE luminescence is observed in the spectral map of this microcrystal, which highlights the efficient collection of charge carriers by the BSF quantum wells. For lower densities of BSFs, the exciton dynamics might involve an initial population of donor-bound exciton states with a subsequent redistribution of the excitons towards the BSFs as detailed in~\cite{Corfdir_apl_2009}.

The presented data emphasize the particular suitability of the investigated sample for the discussion of stacking fault related luminescence lines due to the wide range of emission energies observed and their clear attribution to BSFs and ZB segments in CL images. Apart from the BSFs, the sample is of high structural and optical quality, with the fault-free part of the microcrystals being comparable to state-of-the-art free-standing GaN samples \cite{Dogan_cgd_2011}. Therefore, the obtained results should apply to any samples containing BSFs.

The majority of the peaks in figures~\ref{fig:cl}, \ref{fig:clscan} and \ref{fig:clscan2} are found close to the emission energies associated with the three different types of BSFs. However, the exact peak positions vary and several peaks can be found at intermediate energies. The origins of such shifts are discussed in section~\ref{sec:othershifts}. As a consequence,  we turned to a statistical analysis of peak energies to determine the emission energies related to individual BSFs in \cite{Lahnemann_prb_2012}.

\section{Importance of the spontaneous polarization}\label{sec:psp}

As already indicated, the spontaneous polarization is an essential factor in determining the emission energy of excitons at a specific type of BSF. Together with ZnO and BeO, the group-III nitrides are among the semiconductors exhibiting a significant spontaneous polarization in the WZ phase \cite{Bernardini_prb_1997}, which is a result of the relatively strong ionicity of these crystals \cite{Belabbes_prb_2013}. As a consequence of the higher symmetry of the ZB phase, the spontaneous polarization vanishes for this polytype. The discontinuity of the polarization at the interfaces of BSF QWs induces electric fields, and the QCSE leads to a redshift of the emission energy associated with excitons bound to these QWs. This situation is sketched in figure~\ref{fig:band-scheme}. Clearly, the emission energy depends on the thickness of the QW, i.e.\ on the type of BSF (cf.\ figure~\ref{fig:psp}).
Recently, it has been shown that the change in carrier confinement between the different types of BSFs is small compared to the change in emission energy induced by the QCSE \cite{Lahnemann_prb_2012,Corfdir_jap_2012}. It can therefore be said that the spontaneous polarization is in fact the prime factor determining the emission energy associated with BSFs or ZB segments in a WZ matrix.

For I$_1$ BSFs, Rebane \etal~\cite{Rebane_pssa_1997} get a reasonable agreement between the experimentally observed emission energy and a QW model that does not include the spontaneous polarization (the importance of which was only realized in the same year \cite{Bernardini_prb_1997}). However, already for the I$_2$ BSFs, such a model of a rectangular QW fails. Tischer \etal~\cite{Tischer_prb_2011} speculate on the possible role of a free-electron to localized acceptor transition for the emission energy of I$_2$ BSFs, which would necessitate the formation of acceptors in the immediate vicinity of the BSFs. Instead, taking into account the spontaneous polarization allows for a consistent description of all three types of BSFs and of ZB segments in a QW model.

\subsection{Zinc-blende segments}

%
\begin{figure}[t]
\centering
\includegraphics*[width=8cm]{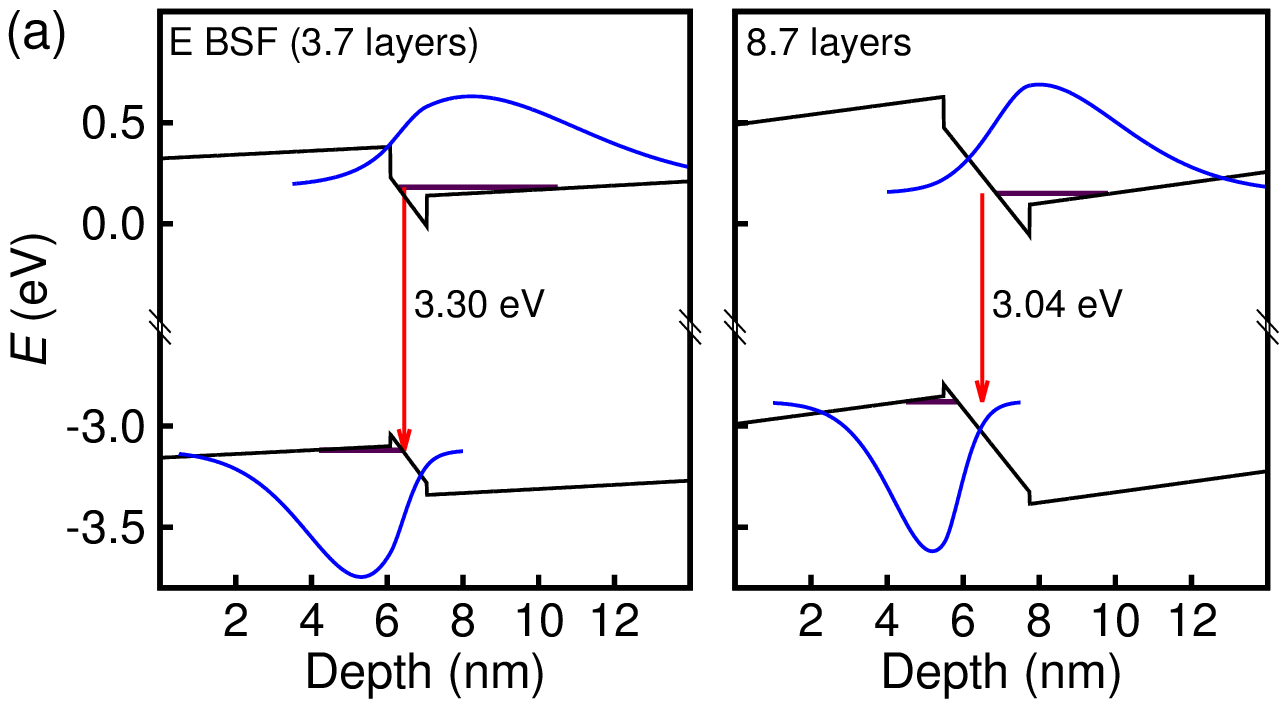}\\[2mm]
\includegraphics*[width=8.2cm]{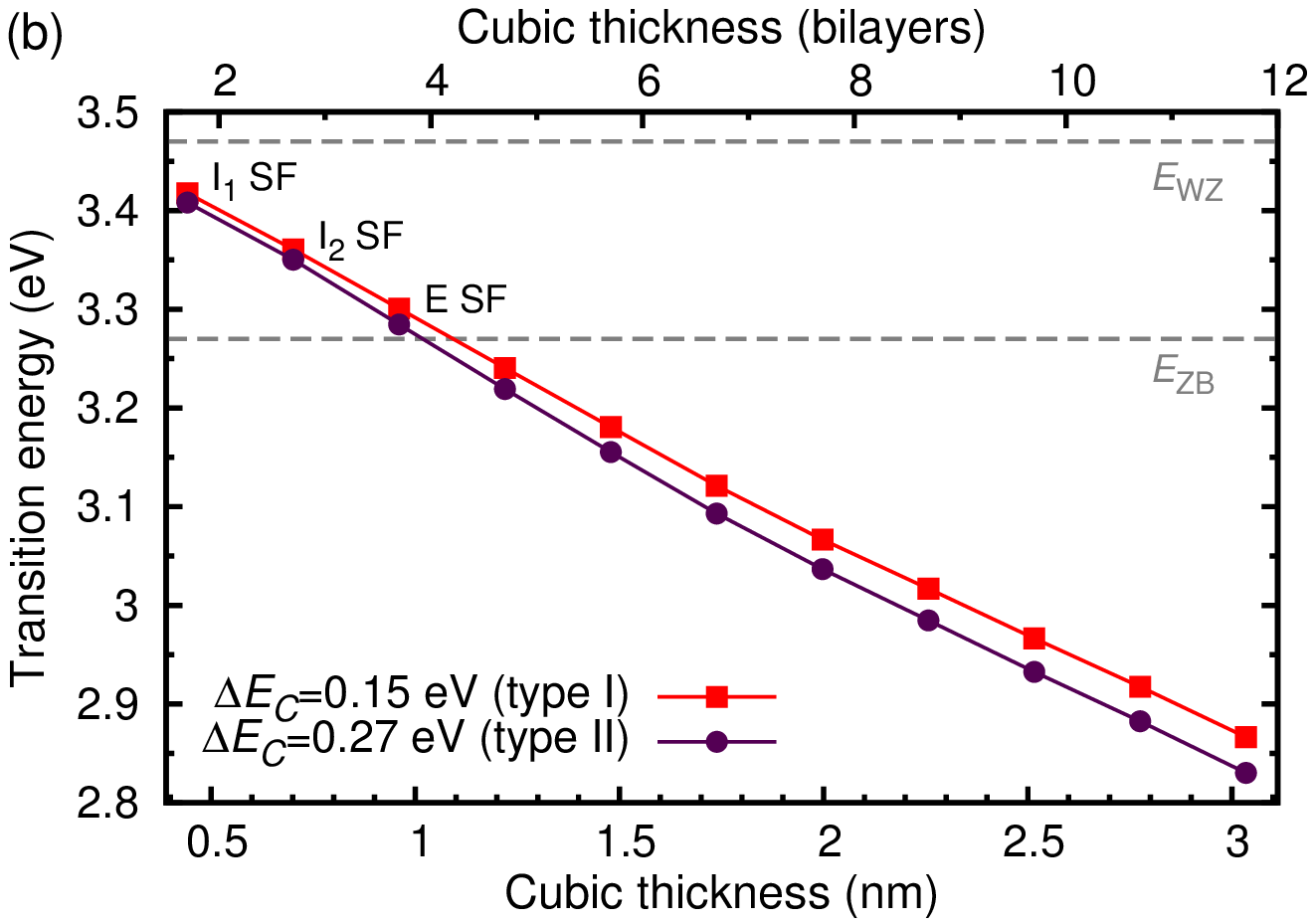}
\caption{\label{fig:thickness}(a) Exemplary band profiles for the E BSF and a ZB segment which is 5 bilayers thicker in a WZ matrix. A type-I band profile is assumed in both cases. The first electron and hole states and their wave functions calculated in the effective-mass approximation are displayed. The vertical arrows indicate the resulting transition energies. (b) Transition energies related to cubic (ZB) QWs of varied thickness. Two different band alignments have been assumed as in \cite{Lahnemann_prb_2012}. The overlap of the electron and hole wave-functions reduces by one order of magnitude over the displayed range. The dashed lines mark the emission energies associated with free excitons in ZB and WZ. All calculations are for polarization fields of 2.5~MV/cm ($P_\mathrm{sp}=-0.022$~C/m$^2$).}
\end{figure}

When the thickness of the ZB segments is increased above that of the extrinsic BSF, the QCSE leads to a further redshift of the emission energy. Excluding longitudinal optical (LO) phonon replicas, the lowest peak energy observed in figure~\ref{fig:clscan2} is 3.0~eV, i.e.\ well below the ZB NBE emission at 3.27~eV. These peaks originate from ZB segments in the WZ crystal, in line with similar observations by Skromme~\etal \cite{Skromme_MaterialsScienceForum_2004} and Jacopin~\etal \cite{Jacopin_jap_2011}. Taking the polarization fields of 2.5~MV/cm ($P_\mathrm{sp}=-0.022$~C/m$^2$) determined in \cite{Lahnemann_prb_2012}, we have calculated the emission energies for cubic segments up to a thickness of 3~nm. The calculations were carried out in the framework of the effective-mass approach through a self-consistent solution of the one-dimensional Poisson and Schr\"odinger equations~\cite{1DPoisson}. We compare a type-I \cite{Belabbes_prb_2011} with a type-II \cite{Stampfl_prb_1998} band alignment, while standard values for the effective masses and dielectric constant of GaN were used~\cite{Vurgaftman_jap_2003}. The results are presented in figure~\ref{fig:thickness}. The band profiles of two exemplary ZB quantum wells---the E BSF and a thicker segment---are shown in figure~\ref{fig:thickness}(a), while the evolution of the emission energy with ZB thickness is given in figure~\ref{fig:thickness}(b). The thickness for the segments is derived from the \emph{effective electronic thickness} for the E BSF that we defined in \cite{Lahnemann_prb_2012}. The calculations are for the single-particle case, but the excitonic transition energies of the ZB and WZ phase have been used in place of the real band gap to account for the exciton binding energy. A negligible change in excitonic binding energy is assumed, which is reasonable as the change should be less than 5~meV across the displayed thickness range according to calculations in \cite{Corfdir_jap_2012}. Note that our calculated transition energies agree with those in \cite{Corfdir_jap_2012} when taking into account the polarization fields and our definition of an effective electronic thickness of the BSF QWs. The different band alignments lead to a relatively small difference in absolute energies that increases slightly with increasing ZB thickness (from 10 to about 35~meV). This result confirms that a type-I band alignment behaves similar to a type-II alignment as the fields resulting from the spontaneous polarization dominate the change in emission energy, and holes are confined in the triangular valence band profile irrespective of the actual band alignment at the interface (cf. figure~\ref{fig:thickness}). Across the thickness range displayed in figure~\ref{fig:thickness}(b), the wave function overlap of the electron and hole decreases by one order of magnitude. When going to ZB segments beyond the E BSF, the emission energy is lower than the excitonic emission energy in ZB GaN. Our lowest observed emission energy at 3.0~eV would correspond to an effective electronic thickness of the ZB segment of 8.7 bilayers, i.e.\ five bilayers more than the E BSF.

Another interesting feature in figure~\ref{fig:clscan} concerns the peak intensities. From the continuous reduction of the wave function overlap with increasing ZB thickness (further spatial separation of electrons and holes) and for a constant contribution from nonradiative recombination, one would expect the emission intensity to continuously decrease from the I$_1$ BSF onwards. Instead, the peak intensities are strongest for intermediate energies (at 3.28~eV in the case of figure~\ref{fig:clscan}) and lower both for the thinnest BSFs as well as for thicker cubic segments. This trend is observed also in other spectral maps (cf.\ figure~\ref{fig:clscan2} as well as figure~3 in \cite{Lahnemann_prb_2012}). The band profiles displayed in figure~\ref{fig:thickness}(a) indicate that the carrier localization provided by the triangular potential in the conduction and valence band as well as the capture efficiency of the QWs increases with the thickness of the ZB segment. These trends are opposite to that of the wave function overlap, and the superposition of both effects can explain the observed development of the peak intensities.

%
\begin{figure}
\centering
\includegraphics*[width=8cm]{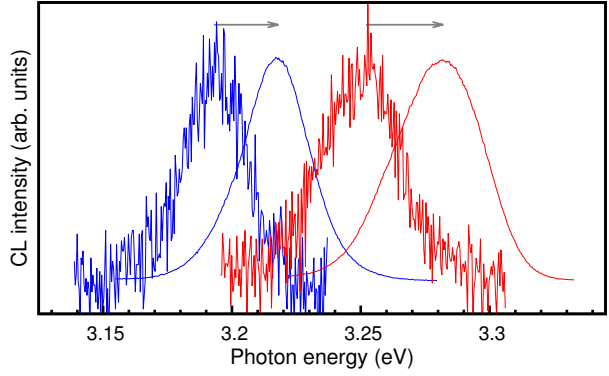}
\caption{\label{fig:shift-spectra}Comparison of spectra recorded with the electron beam away from (noisy) and directly on the ZB segments (smooth). The peaks are blueshifted by a partial screening of the spontaneous polarization field under the increasing excitation. Note that the linewidth and shape of the peaks are not affected by the screening process.}
\end{figure}

\subsection{Excitation-induced screening of the polarization fields} 

%
\begin{figure}
\centering
\includegraphics*[width=8cm]{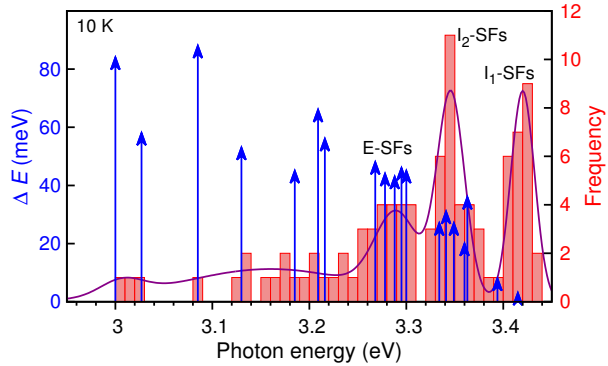}
\caption{\label{fig:shifts}Histogram of (unscreened) emission energies related to BSFs and ZB segments in GaN microcrystals observed in CL spectral maps. The multiple Gaussian fit is a guide to the eye. For energies below that of the E BSF, no clear assignment to a particular ZB thickness is possible. Superimposed is an arrow plot of the maximum blueshift of the peaks  due to screening of the polarization fields observed in the CL maps at the respective energies, which increases from 3~meV at 3.42~eV to 88~meV at 3.09~eV. (Derived from figure 4 in \cite{Lahnemann_prb_2012}.)}
\end{figure}

The previous section has shown that strong polarization fields are necessary to explain the wide range of emission energies observed for ZB segments in WZ GaN. When looking at the CL spectral maps in figure~\ref{fig:clscan} and particularly also figure~\ref{fig:clscan2}, we can see another manifestation of the spontaneous polarization. In these spectral maps, the emission features related to BSFs and ZB segments exhibit a crescent shape. This shape is the result of a partial screening of the polarization fields by the increased carrier concentration and a corresponding blueshift of the optical emission energy when the electron beam approaches the segment~\cite{Lahnemann_prb_2012} which can be as large as 80--90~meV (figure~\ref{fig:clscan2}). To determine unscreened emission energies related to BSFs, we analysed the spectra recorded farthest away from the respective BSF in order to have the lowest possible excitation densities. The shift in emission energy with excitation intensity is direct evidence for the QCSE. A similar shift for BSF emission has been reported from excitation density-dependent PL measurements in \cite{Skromme_MaterialsScienceForum_2004} and \cite{Jacopin_jap_2011} and is also known for stacking faults in SiC~\cite{Juillaguet_apl_2007}. The screening can only be partial, as it is accompanied by an increase in overlap of the electron and hole wave functions and thus a reduction of the radiative lifetimes. As a consequence, the screening of the polarization fields is a self-limiting process, which will saturate at a certain point. Figure~\ref{fig:shift-spectra} exemplifies a comparison of the unscreened and partially screened spectra for two emission peaks related to ZB QWs. The normalized spectra are rigidly blueshifted due to the increased excitation, i.e.\ the linewidth of the peaks is not affected and the symmetric profile is maintained. This property rules out state filling effects as the origin of the blueshift. In figure~\ref{fig:shifts}, we have compiled the maximum shifts in energy we could observe as a result of the screening. The shifts are marked at the unscreened (onset) energy of the respective peaks in the CL spectral maps and are superimposed on a histogram of the unscreened energies of BSF/ZB peaks (compare \cite{Lahnemann_prb_2012}). While the effect is small for the I$_1$ BSFs, it is already significant for the other two types of BSFs and the shift increases with the thickness of the ZB segment. The blueshift of about 45~meV observed for the peaks around 3.29~eV amounts to a screening of one fifth of the value of the polarization fields according to the calculations for varied $P_\mathrm{sp}$ in \cite{Lahnemann_prb_2012}.

As a consequence of this field screening effect, CL spectra integrated over a larger area (e.g.\ figure~\ref{fig:cl}) contain contributions from different degrees of screening and show broadened BSF emission peaks as well as peak energies displaced with respect to the unscreened emission energies. Also for PL measurements, it should be kept in mind that the peak energies are shifted as a function of the excitation density.

\subsection{Time-resolved luminescence spectroscopy}

%
\begin{figure}
\centering
\includegraphics*[width=8cm]{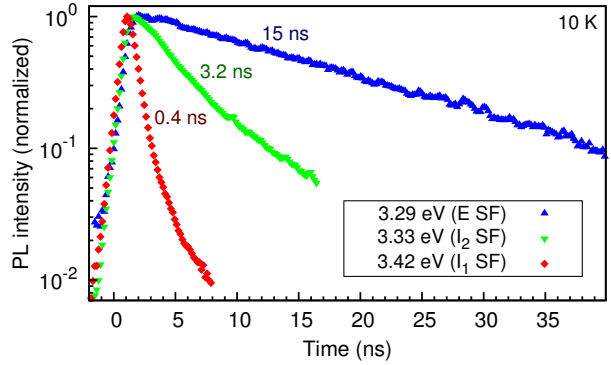}
\caption{\label{fig:trpl}PL transients for emission energies associated with the different types of BSFs measured at low temperatures. The initial decay times are indicated.}
\end{figure}

Another experimental manifestation of the polarization fields and affirmation of the QW model can be found by time-resolved PL measurements. Exemplary transients of BSF emission peaks recorded on one GaN microcrystal at different detection energies are presented in figure~\ref{fig:trpl}. The three transients correspond to the different types of BSFs and thus to different structural thickness. As expected for QWs experiencing polarization fields, where an increased QW thickness should lead to a reduced wave function overlap, the initial decay slows down for lower emission energies. This behaviour also indicates that our transients are dominated by the radiative decay of excitons at the BSFs. Monoexponential fits to the initial decay of the transients in figure~\ref{fig:trpl} give decay times of 0.4~ns at 3.42~eV, 3.2~ns at 3.33~eV and 15~ns at 3.29~eV.

For I$_1$ BSFs, a number of other groups have measured PL or CL transients \cite{Fischer_jcg_1998,Paskov_pssc_2006,Corfdir_apl_2009,
Park_nrl_2011,Corfdir_jap_2009,Furusawa_apl_2013,Korona_jl_2014}. The reported initial decay times vary over an order of magnitude, ranging from 70~ps \cite{Park_nrl_2011} up to 780~ps \cite{Corfdir_jap_2009}. Paskov \etal~\cite{Paskov_pssc_2006} also studied transients at three different emission energies of 3.42, 3.35 and 3.29~eV.  Similar to our results, these authors observe that the initial decay slows down when going to lower emission energies. Therefore, these transients probably correspond to the three types of BSFs as well. At the same time, their study reveals significant differences in the decay rates for samples grown by different methods. The decay times of the higher quality samples grown by hydride vapour phase epitaxy are comparable to our values.

The transient in figure~\ref{fig:trpl} detected at 3.29~eV shows a monoexponential decay over the measurement range, while for the other two transients the decay slows down over time and a multiexponential behaviour is observed. Additional factors such as the electronic coupling of BSFs in a bundle could explain the additional component. Indeed, Corfdir \etal~\cite{Corfdir_apl_2009} observed  a monoexponential decay only for low densities of BSFs and a multiexponential decay for a region of their sample with higher densities of BSFs. This difference was ascribed to a variety of electron-hole overlaps present in the case of electronically coupled bundles of BSFs compared to isolated BSFs. Such a coupling between BSFs, but also a coupling to dopants in their vicinity (see next section) as well as differences in the nonradiative recombination rates depending on the employed growth technique \cite{Paskov_pssc_2006} can explain the range of different decay times observed for I$_1$ BSFs in \cite{Fischer_jcg_1998,Paskov_pssc_2006,Corfdir_apl_2009,Corfdir_jap_2009,
Park_nrl_2011,Furusawa_apl_2013}.

\subsection{Determination of the spontaneous polarization of GaN}

%
\begin{figure}
\centering
\includegraphics*[width=5.5cm]{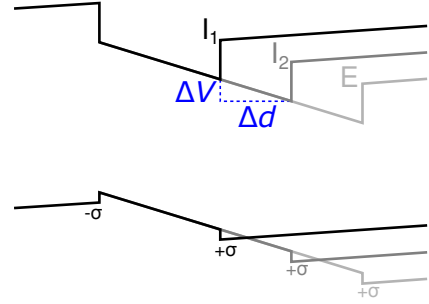}
\caption{\label{fig:psp}Sketch of the band profiles for all three types of BSFs, which illustrates the relationship between changes in the thickness $\Delta d$ of the quantum well and the resulting emission energies.}
\end{figure}

So far, we have emphasized the importance of the spontaneous polarization in determining the emission energies associated with BSFs. In \cite{Lahnemann_prb_2012}, we have turned the tables and used the observed differences in emission energies between the three types of BSFs to determine an experimental value for the spontaneous polarization of WZ GaN. As sketched in figure~\ref{fig:psp}, the change in potential $\Delta V$ when adding a bilayer of thickness $\Delta d$ to the ZB stacking (going from one BSF type to the next) can be associated with the observed change in emission energy. Thus, we employed a plate capacitor model to calculate the polarization charge densities $\sigma$ responsible for the observed shift:
\begin{equation}\label{eq:2}
  \sigma = |P_\mathrm{sp}| = \frac{\Delta V \epsilon \epsilon_0} {\Delta d},
\end{equation}
where $\epsilon$ is the static dielectric constant of GaN and $\epsilon_0$ is the permittivity of free space.
This approach was verified by taking the spontaneous polarization as a free parameter in Poisson-Schr\"odinger calculations, which confirmed that the change of the exciton binding energy and of the confinement in these triangular QW structures are comparably small. Thus, in \cite{Lahnemann_prb_2012}, we derived an experimental value of $P_\mathrm{sp}=-0.022\pm0.007$~C/m$^2$.

\section{Other influences on the emission energies}\label{sec:othershifts}

The \emph{ideal} nature of the BSF quantum wells should lead to both well defined emission energies and very sharp luminescence peaks. Instead, literature reports exhibit a certain range in emission energies---e.g.\ 3.40--3.42~eV for I$_1$ and  3.32--3.36~eV for I$_2$ BSFs. Also, our measurements have shown the presence of BSF-related luminescence at intermediate energies. 

\begin{figure}
\centering
\includegraphics*[width=8cm]{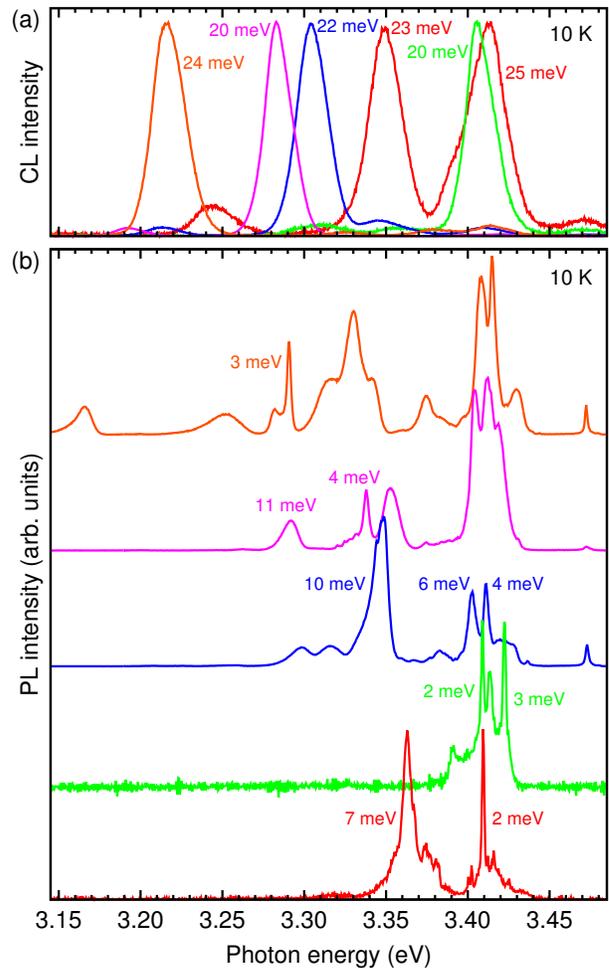}
\caption{\label{fig:pl}(a) Normalized CL spectra of BSF emission extracted from spectral linescans (cf.\ figure~\ref{fig:clscan}) on GaN microcrystals. Contributions from more than one BSF or one bundle in a single spectrum are due to carrier diffusion. The minimum linewidth observable in CL is 20~meV. (b) Normalized low-excitation $\mu$PL spectra of different individual microcrystals containing several BSFs each. Peaks group around 3.42, 3.35 and 3.29~eV, but can be found also at intermediate energies. The linewidths of the individual emission lines are indicated and can be as small as 2~meV for $\mu$PL measurements.}
\end{figure}

Concerning the linewidth (full width at half maximum) of the emission peaks, values above 20~meV are commonly observed in both PL and CL measurements \cite{Paskov_jap_2005,Corfdir_jap_2009,Brandt_prb_2010}. A similar linewidth (20--25~meV) is found in our CL spectra under spot-mode excitation as compiled in figure~\ref{fig:pl}(a). These spectra were recorded at 10~K on the cross sections of two microcrystals. Only BSFs or bundles of these that lie within the scattering range of the electron beam and the diffusion length of the carriers (which for excitons in GaN is around 200~nm  \cite{Ino_apl_2008}) contribute to the spectra. However, CL spectroscopy tends to show a peak broadening both from high excitation densities and the limited spectral resolution of the setup. A narrower linewidth can be found in $\mu$PL measurements. Low-excitation $\mu$PL spectra recorded at 10~K on five microcrystals---detached from the sample and dispersed upside-down on a Si carrier---are displayed in figure~\ref{fig:pl}(b). Not all microcrystals show a contribution from the NBE luminescence in WZ GaN. When observed, the NBE emission is found at its strain-free position. Figure~\ref{fig:pl}(b) demonstrates that a linewidth down to 2~meV can be observed in low excitation $\mu$PL spectra for individual emission lines. However, the linewidth obtained for the donor-bound exciton in the microcrystals investigated here amounts to only 1~meV~\cite{Dogan_cgd_2011} indicating that the BSF emission still suffers some form of broadening. Jacopin \etal~\cite{Jacopin_jap_2011} reported a linewidth down to 0.8~meV on ZB/WZ quantum wells in GaN nanowires, and even lower values ($<0.5$~meV) have been reported for GaAs and InP \cite{Akopian_nl_2010,Graham_prb_2013}.

Several factors can contribute to a shift in energy of BSF-related luminescence. We have already discussed, how the partial screening of the polarization fields can lead to a shift of the observed emission energy as a function of the excitation density. In the following, we will briefly discuss (i) the electronic coupling of bundled BSFs, (ii) the coupling between BSFs and adjacent point defects, (iii) strain induced shifts of the emission energy, and (iv) the intersection of BSFs with ternary quantum wells in non-polar directions. On the one hand, such shifts can lead to the observation of peaks at different energies than expected, which probably is the reason why different reports in the literature exhibit a certain range of emission energies for a specific type of stacking fault. On the other hand, when the shifts in emission energy are on the order of a few meV, the observed peaks can be broadened by the superposition of several peaks.

%

\paragraph{Stacking-fault bundling:}

Basal-plane stacking faults often occur in closely spaced bundles. For distances of only a few atomic layers, this bundling leads to a coupling of the electronic states in neighbouring BSFs and thus to a redshift of the emission energy, as first suggested by Paskov \etal~\cite{Paskov_jap_2005,Paskov_pssc_2006}. In turn, a redistribution of the electric fields for BSF bundles with a slightly larger separation could lead to a blueshift of the emission energy~\cite{Fiorentini_prb_1999}. Recent calculations for I$_1$ BSFs (without spontaneous polarization) show that the coupling plays a role for inter-SF distances of less than 5~nm and can shift the emission energy by up to 30 or 40~meV (for bundles of 2 or 3 BSFs, respectively) in the limiting case of only one layer distance \cite{Corfdir_jap_2012}. By taking into account the spontaneous polarization, the shift is expected to be even larger. Thus, such bundling can lead to emission energies between those of individual BSFs of different types. As bundles of BSFs are visible in transmission electron micrographs of our samples \cite{Lahnemann_prb_2012}, we largely attribute the variation in BSF emission energies observed on these microcrystals and reflected in the statistical analysis of the peak energies to the bundling effect. However, in spectra integrated over a larger area, BSF bundling may also contribute to the broadening of peaks.

\paragraph{Coupling with point defects:}

Point defects in the vicinity of BSFs may affect the emission energy. 
Corfdir \etal \cite{Corfdir_apl_2009,Corfdir_jap_2009,Corfdir_prb_2009} proposed a coupling between a BSF and donor atoms residing in close proximity to this BSF. Such a coupling would influence the emission energy and lead to a localization of excitons within the plane of the BSF. Their calculations for electrons at I$_1$ BSFs suggest that donors with a distance of up to about 10~nm have an influence on the emission energy associated with the BSFs: The difference in electron localization energy with respect to a bare BSF ranges between 9~meV at 10~nm distance to 53~meV when the donor resides exactly on the BSF~\cite{Corfdir_prb_2009}. For the case of acceptors, Khromov \etal \cite{Khromov_prb_2011} discuss the role of Mg impurities located within a few nm of BSFs. However, note that both studies do not include the spontaneous polarization of WZ GaN in their band profiles, which should influence the outcome of such calculations.

For a typical background doping level of $10^{17}$~cm$^{-3}$, the lateral distance of donors coupling to a BSF should be 20~nm on average. Thus, every BSF can potentially couple to a number of donors. The number of populated donor states depends on the excitation density. For low excitation densities, only the deepest donor--BSF states should be populated, which would explain the linewidth of a few meV observable in $\mu$PL.
For higher excitation densities, the statistical distribution of donor--SF distances (assuming that the presence of BSFs does not influence donor incorporation) together with a population of a larger number of these states should essentially lead to a broadening of the peaks associated with individual BSFs~\cite{Corfdir_prb_2009}. The latter situation corresponds to the CL measurements, where a linewidth beyond 20~meV is commonly observed.

Further evidence for an in-plane localization of excitons at I$_1$ BSFs has been reported by several groups in the form of an S-shaped temperature dependence of the peak energy in the range between 10 and 75~K \cite{Paskov_jap_2005,Paskov_pssc_2006,Paskova_procspie_2006,Corfdir_jap_2009,
Korona_jl_2014}. A redistribution of carriers among localized states leads to an initial redshift, followed by a blueshift due to the thermal depopulation of the localized states before the temperature dependence follows that of the free exciton emission \cite{Paskov_jap_2005,Corfdir_jap_2009}. In \cite{Corfdir_jap_2009}, a localization energy of 18~meV was determined. Similar results have been obtained for BSFs in ZnO \cite{Yang_apl_2012}.

\paragraph{Strain-induced shift:}

While the previously discussed effects can contribute both to a shift and a broadening of BSF emission peaks, other factors are expected to induce energetic shifts only. One of them is strain of the semiconductor matrix~\cite{Haberlen_jap_2010}. The band gap and thus the excitonic transition energies in bulk GaN are affected by strain, with compressive strain leading to a blueshift and tensile strain to a redshift of the emission energy~\cite{Perlin_prb_1992,Shan_prb_1996}. Following the QW model of BSFs, the emission energy of excitons bound to these QWs should be shifted accordingly when the band gap of the surrounding matrix is changed.

\paragraph{Intersection of BSFs and QWs:}

For (Al,Ga)N and (In,Ga)N QW structures grown in non-polar directions, it has been reported that the intersection of BSFs with the QWs leads to the formation of one-dimensional quantum wires~\cite{Badcock_apl_2008,Corfdir_jap_2010,Dussaigne_sst_2011,Jonen_apl_2011}. In these quantum wires, the confinement of the QW and the BSF add up at the intersection resulting in an additional peak redshifted with respect to the QW emission. The one-dimensional character of the emission was evidenced by polarization-dependent PL measurements~\cite{Dussaigne_sst_2011}.

\section{BSF-related luminescence at room temperature}\label{sec:rt}

Only few reports include measurements of the emission related to BSFs at elevated sample temperatures~\cite{Salviati_msmc_1997,Mah_jcg_2001,Paskov_jap_2005,Paskov_pssc_2006,Korona_jl_2014}. 
These reports concentrate on the I$_1$ BSFs and tend to see a quenching of this emission line with increasing temperature. However, the temperature-dependent measurements in \cite{Salviati_msmc_1997} and \cite{Paskov_pssc_2006} indicate that the peak might indeed be shifted together with the NBE luminescence, and, as a result of the thermal broadening of the latter, ends up as a mere shoulder of this emission band at room temperature. For GaN nanowires grown at reduced substrate temperatures, the presence of ZB segments can lead to an additional peak at around 3~eV in the room temperature ensemble spectra~\cite{Wolz_nt_2012}.

\begin{figure}
\centering
\includegraphics*[width=8cm]{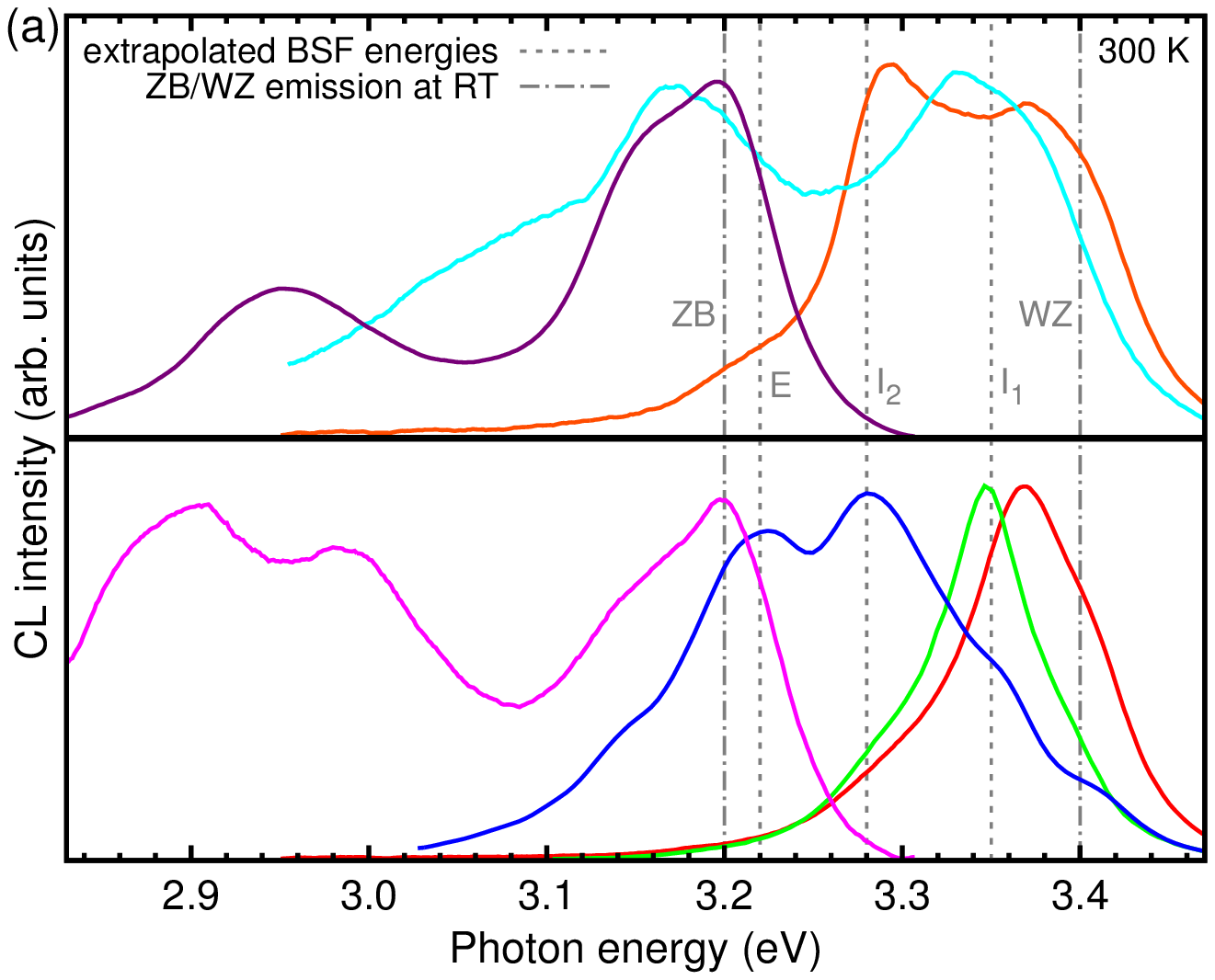}\\[2mm]
\includegraphics*[width=8cm]{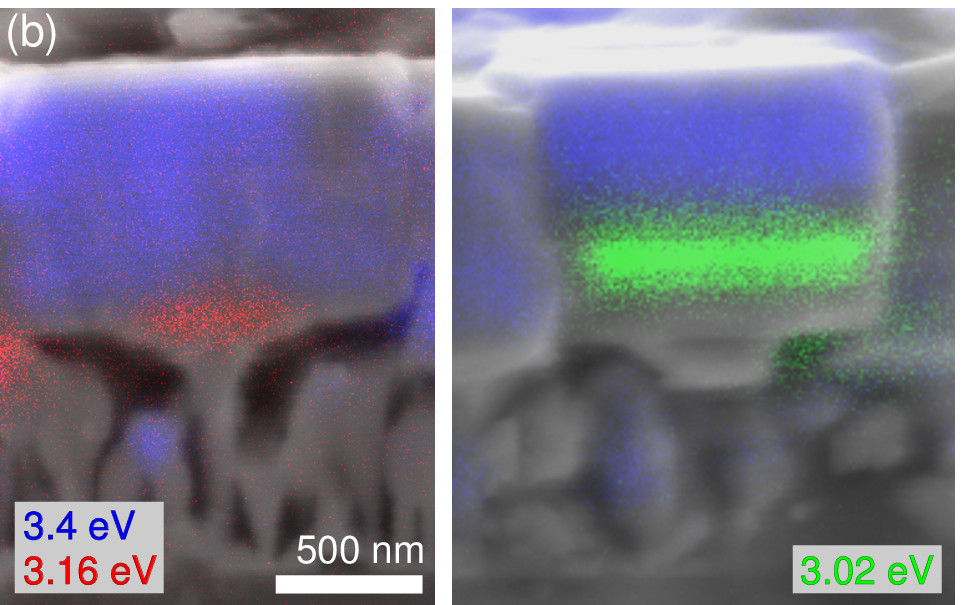}
\caption{\label{fig:rtcl}(a) Normalized room temperature CL spectra recorded with spot-mode excitation at different positions on GaN microcrystals also show BSF-related emission. The indicated emission energies assume that both the band edge and the BSF emission are shifted by 70~meV from their low-temperature values. (b) Monochromatic CL images for two microcrystals recorded at the specified energies superimposed on the corresponding SEM images. The spatial distribution of the emission along the basal plane can be resolved at room temperature for ZB segments.}
\end{figure}

Also for our samples, room temperature transitions at energies both between the WZ and ZB band gaps as well as below that of the ZB phase can be observed in the CL spectra in figure~\ref{fig:rtcl}(a) with peaks in the range between 2.9 and 3.4~eV. Again, spot mode CL measurements provide a way to selectively excite these WZ/ZB quantum wells. However, it is not possible to compile a peak histogram of emission energies and thus to identify the energies related to specific BSF types, because the peak broadening at elevated temperatures prevents the assessment of effects such as the screening of polarization fields. In the QW model, we can assume that the BSF emission shifts by a similar degree as the NBE luminescence when going from 10 to 300~K ($-70$~meV). Then, emission energies of 3.35, 3.28 and 3.22~eV can be expected for the I$_1$, I$_2$ and E BSFs, respectively. These energies are marked in figure~\ref{fig:rtcl}(a) by grey dashed lines, while the NBE emission energies of ZB and WZ GaN at 300~K are marked by dot-dashed lines. In agreement to this expectation, Korona~\etal~\cite{Korona_jl_2014} have recently followed the luminescence of I$_1$ BSFs up to room temperature and determined an emission energy of 3.36~eV. Note that two of the peaks in figure~\ref{fig:rtcl}(a) coincide with the emission energy of ZB GaN, which could indicate the presence of bulk-like (thick) segments of ZB GaN. 

%

This discussion highlights the advantage of the spatially resolved spectral measurements: In spot mode CL spectra, the ZB quantum wells are selectively excited enabling their unambiguous identification, whereas the detection is more difficult in spectra integrated over a larger area \cite{Salviati_msmc_1997,Paskov_pssc_2006} or in maps of monochromatic CL intensities (CL images). For spatially integrated luminescence spectra, the emission related to BSFs might easily be merged into a shoulder of the WZ NBE peak, making it hard to identify peaks associated with BSFs and distinguish them from the phonon replicas. Likewise in CL images, detection windows centred on the BSF emission energies will also contain an additional signal from the phonon replicas of WZ GaN, making it impossible to distinguish the different contributions. Possibly, also the capture of carriers by the BSF QWs is less efficient at elevated temperatures. The identification in CL images on our sample was only possible for ZB segments with a larger localization energy and not for BSFs, e.g.\ for emission energies around 3.16 and 3.04~eV as displayed in figure~\ref{fig:rtcl}(b).

\section{Conclusions}\label{sec:concl}

Since the first identification of luminescence lines attributed to I$_1$ BSFs~\cite{Salviati_msmc_1997,Albrecht_mrssp_1997,Rebane_pssa_1997}, a quite comprehensive picture of the emission properties related to BSFs and ZB segments in GaN has emerged. BSFs in WZ semiconductors act as QWs that collect charge carriers from the surrounding crystal in a very efficient way. Therefore, luminescence spectroscopy provides a sensitive tool to assess the presence of stacking faults in GaN layers without the need for TEM investigations~\cite{Paskov_pssc_2008}. For high BSF densities, the associated luminescence band will even dominate the optical emission spectra at low temperatures~\cite{Paskov_jap_2005,Liu_apl_2005}. In particular, the spatial resolution of CL spectroscopy in an SEM allows for a selective excitation of BSFs as well as the direct imaging of the luminescence distribution characteristic for BSFs. In this topical review, we have given an overview concerning the properties of luminescence associated with BSFs in GaN with an emphasis on the emission energies observed in the literature and on factors which can lead to a shift of these energies.

Particularly, we have shown that the coalescence overgrowth of nanowires is accompanied by the formation of BSFs during the lateral growth. Consequently, it is likely that so far not clearly identified emission energies observed in similar structures are also related to BSFs or cubic segments. Notably, an emission peak at 3.4~eV was observed in \cite{Yang_jcg_2009} and ascribed to the N polarity of the sample, emission lines at 3.32 and 3.42~eV in \cite{Ku_jjap_2010} were attributed to the Y2 and Y6 defect peaks~\cite{Reshchikov_jap_2005}, and a room temperature luminescence peak at 3.19~eV in \cite{Lethy_jap_2012} was attributed to the donor-acceptor pair transition. Also, the emission band at 3.40--3.42 eV related to defects close to the nanowire substrate interface in \cite{Calleja_prb_2000} is likely related to I$_1$ BSFs. Emission bands around 3.35 and 3.29 eV without clearly identified origin are quite commonly observed also in other samples, some of which are known to contain BSFs~\cite{Grandjean_apl_1997,Salviati_pssa_1999,Paskov_jap_2005,Guhne_jap_2007,Zhu_jap_2010}. The luminescence distribution in CL imaging or polarization-dependent measurements might facilitate a distinction from other structural defects such as PSFs.

In this review, we have shown that the properties of emission associated with BSFs can be discussed in a QW model. However, the simple picture of a rectangular QW needs to be extended to include additional factors. Our results emphasize the role of the spontaneous polarization fields for the emission energy of both BSFs and ZB segments in WZ GaN. This role is evidenced by the wide emission range covered by such ZB segments spanning down to energies well below that of the NBE emission in ZB GaN as well as by the partial screening of the polarization fields achieved under high excitation densities. Further evidence for the QW model including polarization fields is found in the PL transients for different BSF-related emission energies. With the parameters established in \cite{Lahnemann_prb_2012}, we have extended our calculations of the BSF-related emission energies to ZB segments of up to 3~nm thickness. Furthermore, using $\mu$PL at low excitation densities, we have confirmed that luminescence from individual BSFs can give rise to emission lines with a linewidth down to a few meV. A broadening of the lines can result from a coupling with point defects (donors) in the vicinity. Other factors leading to a shift of the emission energy associated with BSFs are strain, the screening of polarization fields and a coupling of the QWs in bundles of BSFs.  At room temperature, the efficiency of the carrier localization at the BSFs is reduced, but the related emission band can still be observed. Our CL spectra under local excitation indicate that the emission energy of the BSFs is shifted by a similar degree as that of the NBE emission. For ZB segments, we were able to image the CL distribution along the basal plane even at 300~K.

Many of the considerations presented in this topical review should be applicable in a similar way to other compound semiconductors predominantly crystallizing in the wurtzite structure such as ZnO or SiC, but also to the classic III-V semiconductors, for which wurtzite/zinc-blende polytypism is commonly observed in nanowires. Note in this context that also WZ GaAs exhibits a (though weaker) spontaneous polarization as suggested in \cite{Jahn_prb_2012,Belabbes_prb_2013} and recently observed in \cite{Bauer_apl_2014}.

\appendix

\section{Experimental details}

%
\begin{figure}
\centering
\includegraphics*[width=8.5cm]{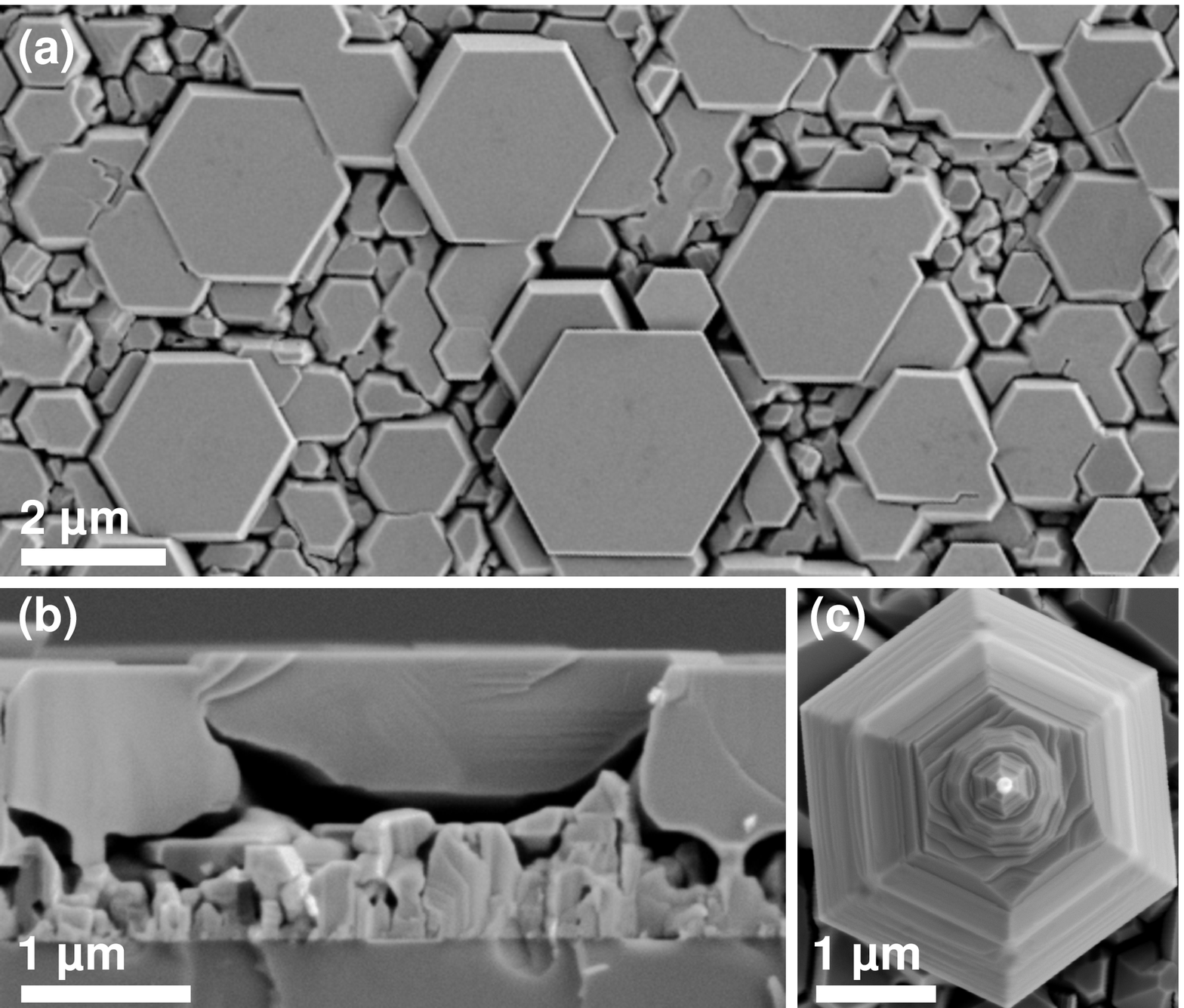}
\caption{\label{fig:sem}SEM images of the GaN microcrystals investigated in this study: (a) Top-view, (b) cleaved cross section, and (c) detached microcrystal lying upside-down and seen from the top (note that the depicted crystal lies on the coalesced layer, while PL measurements were performed on crystals transferred to a foreign substrate).}
\end{figure}

The presented luminescence data were obtained on BSFs and ZB segments formed during the pendeoepitaxial overgrowth of GaN microcrystals on top of GaN nanowires. The growth of GaN by plasma-assisted molecular beam epitaxy under N-rich conditions leads to the self-induced formation of nanowires. To achieve pendeoepitaxial overgrowth of these nanowires, the growth parameters were changed to Ga-rich conditions. This regime promotes lateral growth and eventually leads to the formation of a coalesced GaN layer~\cite{Dogan_cgd_2011}. Scanning electron microscopy (SEM) images of the resulting coalesced layer are shown in figure~\ref{fig:sem}. The layer consists of microcrystals with a diameter of 1--3~$\mu$m embedded in a matrix of smaller grains. The SEM image of a detached and flipped microcrystal in figure~\ref{fig:sem}(c) illustrates that every microcrystal originates from a single nanowire. Probably, the shape transitions during the lateral expansion of the crystal evidenced in figure~\ref{fig:sem}(c) are related to the formation of the BSFs and ZB segments that are the focus of this study.

The emission properties were analyzed using a Gatan Mono-CL3 system. The CL system is mounted to a Zeiss Ultra55 field-emission scanning electron microscope and the spectrometer is equipped with a photomultiplier for monochromatic CL imaging and a charge-coupled device (CCD) detector for spectral mappings of the CL. For cryogenic measurements, the samples can be cooled with liquid He to 10~K. The spectral resolution for the low-temperature CL measurements was set to about 7~meV, and the SEM was operated at an acceleration voltage of 3~kV and a beam current of 2~nA. Additional measurements were performed with a Jobin-Yvon $\mu$PL setup. For these, the beam from the 325~nm line of a Kimmon HeCd laser was focused to a spot diameter of about 3~$\mu$m in a confocal setup. The excitation density was attenuated to about 3~W/cm$^2$. Using a 600 lines/mm grating, the spectral resolution was set to 1~meV and a CCD was employed for detection. Again, the samples were cooled with liquid He. To address single microcrystals, they were dispersed upside-down [cf.\ figure~\ref{fig:sem}(c)] on a Si substrate.
Time-resolved PL was carried out on single dispersed microcrystals using an optical microscope with about 5~$\mu$m spacial resolution and a spectrometer with about 13~meV spectral resolution together with a streak camera. The overall temporal resolution of the setup was 0.8~ns. The sample was excited by laser pulses with 200~fs duration and of a photon energy of 3.531~eV. The energy fluence on the sample surface was 150~$\mu$J/cm$^2$. The repetition rate was set to 4.76~MHz using a pulse picker.

\ack

We would like to thank Pierre Corfdir for valuable discussions, Johannes Zettler for a critical reading of the manuscript, Lutz Geelhaar and Henning Riechert for continuous encouragement and support, as well as Carsten Pf\"uller for assistance with the PL setup.

\bibliography{Laehnemann_SFs_13}

\end{document}